\magnification = \magstep1
\vsize = 22.5 true cm
\hsize = 16 true cm
\baselineskip = 24 true pt
\centerline{\bf {DUALITY IN LIOUVILLE THEORY AS A REDUCED SYMMETRY}} 
\bigskip
\bigskip
\centerline {L. O'Raifeartaigh and V. V. Sreedhar} 
\centerline {School of Theoretical Physics}
\centerline {Dublin Institute for Advanced Studies} 
\centerline {10, Burlington Road, Dublin 4}
\centerline {Ireland}
\bigskip
\bigskip
\centerline {\bf {Abstract}}

The origin of the rather mysterious duality symmetry found in quantum Liouville
theory is investigated by considering the Liouville theory as the reduction of 
a WZW-like theory in which the form of the potential for the Cartan field is 
not fixed {\it a priori}. It is shown that in the classical theory conformal 
invariance places no condition on the form of the potential, but the conformal 
invariance of the classical reduction requires that it be an exponential. 
In contrast, the quantum theory requires that, even before reduction, 
the potential be a sum of two exponentials. The duality of these two 
exponentials is the fore-runner of the Liouville duality. An interpretation  
for the reflection symmetry found in quantum Liouville theory is also obtained 
along similar lines. 
\vfill
\hfill DIAS-STP-99-05\hfil\break
\vfil\eject  
A single real scalar field governed by the so-called Liouville Action [1] 
plays a ubiquitous role in the study of two-dimensional conformal and 
integrable field theories, and is of considerable importance in the context of 
quantum gravity and string theory [2]. Amongst its many unusual properties is 
a particularly intriguing one, namely, the invariance of the quantum theory 
under a duality symmetry which is absent in the classical theory. This duality 
symmetry is responsible for the doubling of a one-dimensional lattice of 
poles (in the parameter space) of the coefficient of the three-point function 
of vertex operators, as explained in detail in [3]. It was also shown there 
that the three-point function, with the correct pole structure, may be derived 
by incorporating the duality into the definition of the path integral for the 
Liouville theory on a sphere.

As is well-known, the classical Liouville theory may be obtained by imposing a 
set of linear first class constraints on the currents of an SL(2, R)
Wess-Zumino-Witten (WZW) model [4]. In the present paper, we investigate 
the origin of the Liouville duality by considering the Liouville theory as the 
reduction of a WZW-like theory in which the form of the potential for the 
Cartan field is not fixed {\it a priori}. It will be shown that the conformal 
invariance of the classical reduction forces this theory to be the SL(2, R)
WZW model, but an essential difference between the classical and quantum 
reductions lies at the heart of the duality symmetry of the quantum Liouville
theory. 

The WZW-like theory we start with is defined by the following Action:
$$S = k\int d^2x\sqrt g ~\lbrack {1\over 2}g^{\mu\nu}(\partial_\mu\phi )
(\partial_\nu\phi ) + g^{\mu\nu}(\partial_\mu a )(\partial_\nu c )
V(\phi )\rbrack\eqno(1)$$ 
In the above equation, $\phi ,~a$ and $c$ are real scalar fields and $V(\phi )$ 
is an arbitrary function of $\phi$. Obviously, (1) describes a conformally 
invariant model at the classical level. Upon the addition of a topological 
term $-ik\epsilon^{\mu\nu} (\partial_\mu a) (\partial_\nu c)V(\phi )$, the 
Action becomes 
$$ S = k\int d^2z~ \lbrack {1\over 2}(\bar\partial\phi )(\partial\phi ) + 
(\bar\partial a)(\partial c)V(\phi )\rbrack\eqno(2)$$
where we have used complex coordinates $z = {x_0 + ix_1\over 2},~\bar z = 
{x_0 - ix_1\over 2}$ in terms of which $\partial = \partial_0 - i\partial_1,~
\bar\partial = \partial_0 + i\partial_1$ and $d^2z = 2idzd\bar z$. 
Note that choosing the topological term with the opposite sign would have 
resulted in $\partial\leftrightarrow\bar\partial$ in (2).
The classical energy-momentum tensor, $T^{\alpha\beta} \equiv 
-{1\over\sqrt g} {\delta S\over\delta g_{\alpha\beta}}$, is given by 
$$\eqalign{ T^{\alpha\beta} =  {k\over 2}& g^{\alpha\beta}g^{\mu\nu}\lbrack 
{1\over 2}(\partial_\mu\phi ) (\partial_\nu\phi ) + (\partial_\mu a)
(\partial_\nu c )V(\phi )\rbrack \cr& - kg^{\mu\alpha}g^{\beta\nu}\lbrack 
{1\over 2}(\partial_\mu\phi ) (\partial_\nu\phi ) + (\partial_\mu a)
(\partial_\nu c )V(\phi )\rbrack }\eqno(3)$$
As usual, it is traceless, symmetric, and conserved, and does not receive  
any contribution from the topological term as the latter is independent of the 
metric. 

Note that for $V(\phi ) = e^{-\phi} $, Eq. (2) simplifies to the SL(2, R) 
WZW Action, with $a$ and $c$ locally parametrising the nilpotent subgroups, 
and $\phi$ the abelian subgroup, respectively, in a Gauss decomposition of the 
group-valued WZW field.  In the following, however, we shall allow $V(\phi )$ 
to be arbitrary to begin with, and examine the restrictions imposed on its 
form, by requiring the Action (2) to be amenable to a conformally invariant 
reduction. It will be shown that for this requirement to be satisfied, 
$V(\phi )$ is governed by a second order functional differential equation 
whose two solutions are dual to each other. In the classical limit only one of 
the two solutions of this equation has a non-trivial ($\phi$-dependent) value 
which corresponds to the standard SL(2, R) WZW case.   

We shall first present the analysis of conformal invariance in the classical 
case. The expressions for the momenta $\pi$ conjugate to the fields 
$\phi ,~a,~c$ follow from the Action (2) and are given by 
$$\pi_\phi = k\dot\phi ,~~~\pi_a = kV(\phi )\partial c,
~~~{\hbox {and}}~~~\pi_c = kV(\phi )\bar\partial a\eqno(4)$$ 
The non-vanishing Poisson brackets, evaluated at equal time, are given by 
$$\{\phi (x, t), \pi_\phi (y, t)\} = \{ a(x, t), \pi_a (y, t)\} = \{ c(x, t), 
\pi_c (y, t)\} = \delta (x - y) \eqno(5)$$ 
The Virasoro generator $T$ is given by 
$$T = \int dz~\epsilon (z)\lbrack -{k\over 2}(\partial\phi)^2 - k\pi_a\pi_c 
V^{-1} + 2i\pi_a(\partial_1a ) \rbrack (z)\eqno (6)$$
The other Virasoro generator $\bar T$, may be obtained by making the exchanges
$a\leftrightarrow c,~\partial\leftrightarrow\bar\partial,~\epsilon\rightarrow
\bar\epsilon$.  As usual, the Poisson bracket of the two Virasoros vanishes, 
and it suffices to focus our attention on one Virasoro.    

The conformal transformation of a field ${\cal O}$, generated by $T_\epsilon$, 
is given by $\delta_\epsilon{\cal O} \equiv \{T_\epsilon , {\cal O}\}$.  
Hence the fields have the following conformal variations: 
$$\delta_\epsilon a = \epsilon\partial a,~~~\delta_\epsilon c = 
\epsilon\partial c,~~~\delta_\epsilon\phi =\epsilon\partial\phi ,~~~ 
\delta_\epsilon V(\phi) = \epsilon\partial V(\phi ) \eqno(7)$$ 
$$\delta_\epsilon \pi_a  = \epsilon\partial \pi_a + (\partial \epsilon )
\pi_a,~~~\delta_\epsilon \pi_c = 0\eqno(8)$$ 
The momentum $\pi_\phi$ does not transform like a conformal primary field:  
$$\delta_\epsilon \pi_\phi = \epsilon\partial\pi_\phi + (\partial
\epsilon ){k\over 2}\partial\phi$$ 
This is expected because $\pi_\phi = (k\dot\phi )$ is a mixture of $\partial$
and $\bar\partial$ derivatives of a scalar field. As can be seen from the 
following equation, however, $\partial\phi$ transforms like a conformal 
primary field:  
$$\delta_\epsilon \partial\phi = \epsilon\partial\partial\phi + (\partial
\epsilon )\partial\phi\eqno(9)$$ 
It is also easy to check that 
$$\delta_\epsilon T  = \epsilon\partial T + 2(\partial \epsilon )T\eqno(10)$$ 
It follows from the above equations that classically $a,~c,~\phi ,~\pi_c$ have 
conformal weights zero; $\pi_a,~\partial\phi$ have weights one, and $T$ has a 
weight two -- as expected. The weights with respect to the other Virasoro are 
given by the same numbers if the exchanges $a \leftrightarrow c$, 
$\epsilon \leftrightarrow \bar\epsilon $, and $\partial \leftrightarrow \bar
\partial$ are done.  

At this juncture it is worth noting that the Virasoro (6) is unique if we 
insist on keeping $V(\phi )$ arbitrary. However, if we are interested in 
a special class of theories for which  $V$ is restricted to be of the form 
$$ V = e^{\lambda\phi }\eqno(11)$$
the Virasoro is no longer unique.\footnote{*} {Note that by scaling the fields 
$\phi ,~a,~{\hbox{and}}~c$, appropriately, (2) can be seen to represent 
the SL(2, R) WZW model for this choice of $V$.} 
This is because it is not necessary that $\pi_a$ and $\pi_c$ should have 
conformal weights (1,0) and (0,1) respectively, for the Action (2) to be 
conformally invariant; it is only necessary that the combination 
$\pi_a\pi_c V^{-1}(\phi )$ has a weight (1,1). This freedom is expressible in 
terms of improvement terms which, when added to the standard Virasoro (6), 
redistribute the weights of the various fields in a way which preserves the 
weight of the above combination of the fields. It is straightforward to see 
that the improvement term $t$, for the Virasoro $T$, is given by 
$$t_\epsilon = -\alpha\int dz~(\partial\epsilon )\lbrack k\partial\phi -  
\lambda a\pi_a \rbrack (z) \eqno(12)$$
where $\alpha$ is an arbitrary parameter. Clearly, with respect to the 
full Virasoro $T+t$, $\pi_a$ behaves like a primary field of weight $(1+\lambda
\alpha , 0)$. It is also easy to see that the field $e^{-\lambda\phi}$ 
transforms as follows under the action of the full Virasoro $T + t$: 
$$\delta_\epsilon e^{-\lambda\phi} = -\lambda\alpha (\partial\epsilon )
e^{-\lambda\phi} + \epsilon \partial e^{-\lambda\phi}\eqno(13)$$      
Thus the combination $\pi_a\pi_cV^{-1}$ has weight one with respect to the 
full Virasoro $T+t$. Note that the ratio of the coefficients of the two terms 
in (12) is fixed by the parameter $\lambda$ in the potential. It is 
straightforward to see that this ratio is also equal to the corresponding 
ratio in the improvement terms $\bar t$ for the other Virasoro $\bar T$. 
Further, the requirement that the Poisson bracket of the two Virasoros is zero 
implies that the improvement terms $t$ and $\bar t$ have the same overall 
coefficient $\alpha$ relative to $T$ and $\bar T$. It may also be 
noted that for the theory to remain conformally invariant, it is not possible 
to have only $\lambda$ to be zero. If $\lambda$ is zero, then $\alpha$ has 
to be necessarily zero.    

It is worth mentioning that there is a good reason for considering the 
improved Virasoro $T+t$, and the concomitant freedom in redistributing the 
weights of the various fields in the theory, namely, that it is necessary for  
the conformal reduction of the SL(2, R) WZW theory to the Liouville theory. The 
constraints that accomplish this reduction are 
$$ \pi_a = m_a~~~{\hbox {and}}~~~ \pi_c = m_c\eqno(14)$$
where $m_a$ and $m_c$ are constants.  Note that unless the improvement terms
are present with $\lambda\alpha  = -1$, it is not possible to impose the 
above constraints in a conformally invariant manner since $\pi_a$ and $\pi_c$ 
have non-zero conformal dimensions. Thus one is forced to use the full 
Virasoro $T+ t$ with $\lambda\alpha = -1$. Indeed, the improvement term with 
$\lambda\alpha = -1$ automatically provides the usual improvement terms of 
the Liouville theory upon reduction. Having thus motivated the need for 
the improvement terms, it may be noted that the argument can be reversed:
Given the full Virasoro $T + t$ and the Action (2), conformal invariance 
requires $V$ to be of the form (11). 

The situation in the quantum theory is remarkably different from that in the 
classical theory. The entire analysis leading up to Eq. (10) can be carried 
over to the quantum theory in a relatively straightforward way by replacing 
Poisson brackets with commutators and the appropriate factors of $i\hbar$, as 
long as care is exercised in taking the ordering of the operators. The 
commutator of $\pi_a$ with the Virasoro can be calculated without ado as the 
Virasoro is linear in $a$, the conjugate field. As expected, $\pi_a$ remains a 
conformal primary field of weight (1, 0). The commutator of $V^{-1}$ with the 
Virasoro is trickier because $T$ is quadratic in $\pi_\phi$. A short 
calculation shows that 
$$\delta_\epsilon V^{-1}\equiv\lbrack T_\epsilon ,~ V^{-1}\rbrack_-
= (\partial\epsilon ) \Bigl( {-i\hbar^2\over 2k} {\delta^2V^{-1}\over \delta
\phi^2}  \Bigr) + i\hbar \epsilon (\partial V^{-1})\eqno(15)$$
where we have used $-\partial\delta (z - z') = i[\delta (z-z')]^2$ [5].
Thus, at the quantum mechanical level, $V^{-1}$ ceases to behave like a 
conformal scalar for general $V$, with respect to $T_\epsilon$. However, as in 
the classical case, if one adds improvement terms, it is sufficient to require 
that the combination $\pi_a\pi_cV^{-1}(\phi )$ has a weight (1,1). It is easy 
to check then that with respect to the full Virasoro $T + t$, $\pi_a$ behaves 
like a primary field of weight $(1+ \lambda\alpha , 0)$ even at 
the quantum level. The conformal variation of $V^{-1}$ with respect to the 
full quantum Virasoro is  
$$\delta_\epsilon V^{-1}\equiv\lbrack T_\epsilon + t_\epsilon,~ V^{-1}\rbrack_-
= (\partial\epsilon ) \Bigl( {-i\hbar^2\over 2k} {\delta^2V^{-1}\over \delta
\phi^2} + i\hbar \alpha {\delta V^{-1}\over \delta \phi} \Bigr) + 
i\hbar \epsilon (\partial V^{-1})\eqno(16)$$
The necessary condition for $V^{-1}$ to have the correct conformal properties, 
can be read off from the above equation to be 
$${-\hbar\over 2k\alpha}{\delta^2V^{-1}\over\delta\phi^2} + {\delta V^{-1}\over
\delta\phi} = -\lambda V^{-1} \eqno(17) $$     
Note that the above equation is a quadratic equation and has  two solutions 
while its classical counterpart obtained by setting $\hbar = 0$ is a first 
order equation which leads to $V = e^{\lambda\phi}$. Thus the quantum theory 
is slightly less stringent about the form of $V$, than the classical theory,  
with respect to the demands imposed by conformal invariance. The general 
solution for $V^{-1}$ takes the form 
$$ V^{-1} = e^{\omega_+\phi} + \mu e^{\omega_-\phi}\eqno(18)$$
where 
$$\omega_\pm = {\alpha k\over\hbar} \mp {\alpha k\over\hbar}
\sqrt{ 1 + {2\hbar\lambda\over \alpha k}}
\eqno(19)$$
and $\mu$ is a constant. Note that $\omega_+\omega_- = -{2k\lambda\alpha\over
\hbar}$~ --- showing that the two solutions are dual to each other. Also note 
that in the classical limit, $\hbar \rightarrow 0,~~\omega_+\rightarrow 
-\lambda~~{\hbox {and}}~~ \omega_- \rightarrow \infty$. In this 
limit the duality disappears and only (11) survives. 

In passing we mention that the above results can be readily applied to 
work out the transformation properties of the vertex operators $U(\phi ) = 
e^{2\gamma\phi}$  of the field $\phi$. It follows from Eq. (16) that 
$$\delta_\epsilon (e^{2\gamma})\equiv\lbrack T_\epsilon + t_\epsilon,~e^{2
\gamma} \rbrack_- = (\partial\epsilon ) (i\hbar ) \Bigl({2\gamma\over k}
(k\alpha- \hbar\gamma )\Bigr)e^{2\alpha} + i\hbar\epsilon (\partial e^{2\gamma})
\eqno(20)$$
Thus the conformal weight of the vertex operator is given by  
$$\Delta (e^{2\gamma}) =  {2\gamma\over k} (k\alpha - \hbar\gamma )\eqno(21)$$ 
which is manifestly symmetric under the exchange $\gamma \rightarrow k\alpha -
\gamma$ in units of $\hbar = 1$. This is the reflection symmetry of the 
conformal weight of a vertex operator in the quantum theory. 
 
Substituting the solution for $V$ from Eqs. (18) and (19) in Eq. (2), we have 
$$ S = -\int d^2z~ \lbrack {1\over 2}(\bar\partial\phi )(\partial\phi ) + 
(\bar\partial a)(\partial c) (e^{\omega_+\phi} + \mu e^{\omega_-\phi })^{-1}
\rbrack\eqno(22)$$ 
What we have shown is that the above highly non-linear Action is conformally 
invariant with respect to the full Virasoro $T+t$ in the quantum theory, and 
reduces to the standard SL(2, R) WZW case in the classical limit. It easily 
follows that the constraints (14) reduce (22) to  
$$ S = -\int d^2z~ \lbrack {1\over 2}(\bar\partial\phi )(\partial\phi ) + 
m_am_c (e^{\omega_+\phi} + \mu e^{\omega_-\phi }) \rbrack\eqno(23)$$ 
This is precisely the theory proposed in [3] to explain the duality symmetry 
of quantum Liouville theory and the pole structure of its three-point function.

To summarise the results of this paper, we have shown how the duality symmetry 
that appears rather mysteriously in quantum Liouville theory may be interpreted
as a reduced symmetry. Thus the origin of this symmetry can be traced to the 
fact that the conditions imposed by conformal invariance on the structure of 
the potential in the Liouville theory are different in the classical and 
quantum reductions, the latter being less stringent than the former. This is 
because the $\hbar$-dependent operator ordering effects in quantum theory 
produce a second order functional differential equation for the potential, 
whose solutions are dual to each other, while the absence of such effects in 
the classical theory produces a first order functional differential equation 
whose unique solution does not allow for the possibility of a duality symmetry.
A natural interpretation for the reflection symmetry in quantum Liouville 
theory was also obtained along these lines. It would be interesting to see 
if the results of this paper can be generalised to discuss the duality and 
reflection invariance of Toda theories as reduced symmetries.  
\vfil\eject
\centerline {\bf {REFERENCES}}
\bigskip
\item {1. } H. Poincare, J. Math. Pure App. {\bf 5} {\it se 4} (1898) 157;
N. Seiberg, Notes on Quantum Liouville Theory and Quantum Gravity, in 
"Random Surfaces and Quantum Gravity", ed. O. Alvarez, E. Marinari, 
and P. Windey, Plenum Press, 1990.
\item {2. } A. Polyakov, Phys. Lett. {\bf B103} (1981) 207; T. Curtright 
and C. Thorn, Phys. Rev. Lett. {\bf 48} (1982) 1309; E. Braaten, T. Curtright 
and C. Thorn, Phys. Lett {\bf 118} (1982) 115; Ann. Phys. {\bf 147} (1983) 365;
J.-L. Gervais and A. Neveu, Nuc. Phys. {\bf B238} (1984) 125; {\bf B238}
(1984) 396; {\bf 257}[FS14] (1985) 59; E. D'Hoker and R. Jackiw, Phys. Rev. 
{\bf D26} (1982) 3517.
\item{3. } H. Dorn and H.-J. Otto, Phys. Lett {\bf B291} (1992) 39; Nuc. Phys.
{\bf B429} (1994) 375; A. and Al. Zamolodchikov, Nuc. Phys. {\bf B477}(1996)
577; L. O'Raifeartaigh, J. M. Pawlowski, and V. V. Sreedhar, Duality 
in Quantum Liouville Theory, {\it hep-th/9811090}; {\it to appear in 
Annals of Physics.} 
\item {4. } F. A. Bais, T. Tjin and P. Van Driel, Nuc. Phys. {\bf B 357} (1991) 
632; V. A. Fateev and S. L. Lukyanov, Int. J. Mod. Phys. {\bf A3} (1988) 507;
S. L. Lukyanov, Funct. Anal. Appl. {\bf 22} (1989) 255; P. Forg\'acs, A. Wipf,
J. Balog, L. Feh\'er, and L. O'Raifeartaigh, Phys. Lett. {\bf B227} (1989)214;
J. Balog, L. Feh\'er, L. O'Raifeartaigh, P. Forg\'acs and A. Wipf, Ann. Phys. 
{\bf 203} (1990) 76; Phys. Lett {\bf B244}(1990)435; L. Feh\'er, 
L. O'Raifaertaigh, P. Ruelle, I. Tsutsui and A. Wipf, Phys. Rep. {\bf  222}
No. 1, (1992)1; P. Bouwknegt and K. Schoutens Eds. W Symmetry (Advanced 
series in mathematical physics, 22) World Scientific, Singapore (1995);   
L. O'Raifeartaigh and V. V. Sreedhar, Nuc. Phys. {\bf B520}
(1998) 513; Phys. Lett. {\bf B425} (1998) 291. 
\item {5. } C. Ford and L. O'Raifeartaigh, Nuc. Phys. {\bf B460} (1996) 203. 
\vfil\eject\end